# Radiation of a variable charge flying into medium

**Xingzhen Fan and Andrey V. Tyukhtin**

Saint Petersburg State University
7/9 Universitetskaya nab., St. Petersburg, 199034 Russia

E-mail: a.tyuhtin@spbu.ru

## Annotation

We study an electromagnetic radiation of a small bunch having a variable charge value and crossing a flat interface between two media. Both media are homogeneous, stationary, and isotropic. They may have frequency dispersion but no spatial dispersion. Cherenkov radiation can be generated in the second medium only. It is assumed that the bunch charge value decreases exponentially starting from some time moment after the charge enters the second medium. This means that we take into account the scatter of the particle path lengths, connected with a statistical nature of energy losses of particles. It is taken into account that the filamentous "trace" consisting of immobile charges is formed in the second medium.

We obtain the general solution of the problem which is the sum of the forced field, i.e. the charge field in the unbounded medium, and the free field connected with the influence of the boundary. The asymptotic study for the wave zone is carried out. We obtain expressions for the spherical wave as well as for the cylindrical wave generated in the second medium if the charge velocity is sufficiently high. The spherical wave is radically different from the usual transition radiation, since it consists of two parts: the transition radiation wave and the wave generated due to the bunch decay and the formation of its "trace".

We describe the main properties of radiation. If the process of the bunch decay starts from the boundary between media, then the angular distribution of the radiation energy has the single maximum in each from two regions. In the second medium, the radiation, as a rule, is greater than in the vacuum area, and this difference increases with increasing the charge velocity. If the charge velocity is higher than the speed of light in the second medium, then, along with appearance of Cherenkov radiation, the properties of spherical wave change radically. In particular, the main maximum of the angular distribution of the radiation energy increases sharply. If the distance from the charge entry point to the region of the bunch decay is sufficiently large, then the complex interference pattern with many extrema arises due to the summation of different spherical waves.

## Introduction

Charged particles moving in the medium generate electromagnetic radiation in the very wide frequency range, from radio waves to gamma rays. The study of different types of radiation requires different approaches. In the field of x-ray and gamma radiation, it is usually necessary to apply the methods of quantum physics, while at the lower frequencies, radiation is satisfactorily described within the framework of classical electrodynamics. In this study, we focus on the radiation of a particle bunch as a whole, i.e. we consider wavelengths exceeding the size of typical beams. Thus, we are talking about the frequencies from gigahertz to tens of terahertz where the macroscopic electrodynamics apparatus is applicable.

It should be noted that the series of monographs and the huge number of journal papers are devoted to the problems of electromagnetic radiation of moving charges in material media (see, for instance, [1–9]). Usually, in such problems, it is assumed that the particle bunch has some constant velocity, and the value of its charge does not change during the motion. An exception is problems of radiation in the dielectric waveguide structures where the bunch moves in a vacuum channel: in such a situation, it is often taken into account the dynamics of the bunch associated with interaction of particles of bunch [10, 11].

However, if the bunch moves in the medium then its particles interact with the medium particles, and it leads to certain changes of the bunch. In more or less dense media, this interaction is usually the main factor determining the bunch evolution. Various variants of this evolution have been described in many monographs and articles (see [12–17] and references therein).

Depending on the particle mass and velocity, as well the medium density and other factors, it is possible both a rapid deviation of particles from a rectilinear trajectory, leading to the beam scattering, and an almost rectilinear motion of the bunch. The last variant is typical for bunches of heavy particles (protons and ions). It should also be noted that these particles usually lose the most part of their energy in a relatively short section of the trajectory, known as the "Bragg peak" [12-17]. Because of this property, bunches of heavy particles find wide application in medicine (proton and ion therapy).

The regularities characterizing the passage of the charged particles bunches through the matter are quite complex and depend on many factors (energy, mass and charge of particles, the matter density, etc.) [12–17]. In this paper, we restrict ourselves to the following model. We assume that each particle of the bunch moves at some constant speed until some moment when it stops instantly. The speeds of all particles until the stopping moment are the same. However, the moments of stopping for different particles are different. This means that we take into account the scatter of the particle path lengths, connected with the statistical nature of energy losses of particles [12, 14].

Thus, the number of bunch particles decreases: the moving particles turn into non-moving ones. We will assume that certain fraction of particles stops per the unit length of the bunch path. This means that the bunch charge decreases exponentially with time. This model of the bunch evolution is one of the simplest models. It allows performing the complete analytical calculation of the generated radiation and describing the main physical effects.

Previously, we analyzed the radiation of the described bunch with variable charge in a homogeneous infinite medium [18]. For applications, it is more important the case when the bunch flies from the first medium (whose properties are usually close to the vacuum) into the second medium having some relatively high refractive index. We will focus on this problem in the present work.

We assume that the bunch charge does not change in the first medium due to the low density of it. In the second medium, the charge value remains practically constant until certain time moment, starting from which the “bunch decay” occurs. Of course, the particular case of coincidence of the charge entry moment and the moment of the bunch decay beginning can be considered also within the framework of this model. In the situation described, the generated wave field is a combination of three types of radiation: transition radiation, radiation due to the charge decreasing process, and Cherenkov radiation (if the bunch velocity is sufficiently high).

## 1. Formulation of the problem

We will analyze a radiation with wavelengths significantly exceeding the particle bunch size. In this case, the charge can be considered as pointed. The charge speed $\vec{v} = c\vec{\beta}$ is assumed to be constant, and the charge value $q$ is variable. In order to satisfy the charge conservation law (continuity equation), certain "additional" non-moving source with the charge density $\rho_1$ (the "trace" of the initial bunch in the medium) has to be introduced. Combining the axis $z$ with the line of the charge motion, we can write the total charge ( $\rho_\Sigma$ ) and current ( $\vec{j}_\Sigma$ ) densities in the following form:

$$
\begin{aligned}
&\rho_\Sigma = \rho + \rho_1, \\
&\rho = q(t)\delta(x,y,z-vt), \\
&\rho_1 = -\left.\frac{dq(t')}{vdt'}\right|_{t'=z/v} \delta(x,y)\Theta(vt-z) = -\frac{dq(z/v)}{dz}\delta(x,y)\Theta(vt-z), \\
&\vec{j}_\Sigma = \vec{j} = v\rho\vec{e}_z,
\end{aligned}
\tag{1.1}
$$

where $\Theta(\zeta) = 1$ for $\zeta > 0$, and $\Theta(\zeta) = 0$ for $\zeta < 0$. Substituting these expressions into the continuity equation $\operatorname{div}\vec{j}_\Sigma + \frac{\partial \rho_\Sigma}{\partial t} = 0$, it is easy to verify that it turns into an identity.

The formation of the "trace" means that the bunch particles stop due to the interaction with the medium particles, i. e. they turn from moving to immobile (therefore $\vec{j}_1 = 0$). From the point of view of macroscopic electrodynamics, the detailed description of this process is of no importance. For example, this can be the recombination of the bunch electrons with ions (if the medium is a plasma), the stopping of particles due to collisions with neutral molecules, etc. For us, only the fact of the formation of the filamentous "additional" charge is important.

It is assumed that the particle bunch moves perpendicularly to the interface between two media, flying from the medium 1 ( $z < 0$ ) into the medium 2 ( $z > 0$ ) (Fig. 1.). Both media are considered to be homogeneous, linear, stationary, isotropic, and do not have any spatial dispersion (frequency dispersion can take place). The media are characterized by permittivities $\varepsilon_{1,2}$ and permeabilities $\mu_{1,2}$. Accordingly, the wave numbers are equal to $k_{1,2} = k_0 n_{1,2}$, where $k_0 = \omega/c$ is the wave number in vacuum, $n_{1,2} = \sqrt{\varepsilon_{1,2}\mu_{1,2}}$ are the complex refractive indices of the media. Initially, we will assume that the media have some conductivity, i.e. $\operatorname{Im}\varepsilon_{1,2} > 0$ at $\omega > 0$. Ultimately, we will be interested in the case when the conductivity is negligible, and absorption of radiation in the medium is insignificant.

We will consider here only ordinary ("right") media, i.e. we assume that the real values of the refractive indices take place only for real positive values of permittivity and permeability ( $\varepsilon_{1,2} > 0$ and $\mu_{1,2} > 0$ ). The case of a "left" medium (where both constants are negative in the same frequency range) can be considered similarly. Here we exclude this case from consideration for the sake of the relative compactness of formulas. Note that radiation generated by flying of a constant charge into a "left" medium was studied in [19, 20].

We assume that the bunch charge is constant ( $q = q_0$ ) up to the moment $t = t_0 > 0$, i.e. in the medium 1 and, possibly, on some part of the trajectory in the medium 2. At $t > t_0$, the charge value decreases. Let us assume that for any small time interval $dt$ the bunch loses the same fraction of the charge $dq$, that is $dq/dt = q/\tau$, where $\tau = const$. Solving this equation we find

$$q(t) = \begin{cases} q_0 \text{ for } t < t_0, \\ q_0 \exp\left(-\left(t-t_0\right)/\tau\right) \text{ for } t \geq t_0. \end{cases} \tag{1.2}$$

In this case, the motionless "trace" of charges has the density

$$\rho_1 = \frac{q_0}{v\tau} \exp\left(-\frac{z-vt_0}{v\tau}\right) \delta(x,y) \Theta(vt-z) \Theta(z-vt_0) \tag{1.3}$$

where $\Theta(\xi) = 1$ for $\xi > 0$ and $\Theta(\xi) = 0$ for $\xi < 0$.

## 2. Forced field

The total charge field in both media can be represented in the form

$$\vec{E}^{(1,2)} = \vec{E}^{\mathrm{q}(1,2)} + \vec{E}^{\mathrm{b}(1,2)}, \quad \vec{H}^{(1,2)} = \vec{H}^{\mathrm{q}(1,2)} + \vec{H}^{\mathrm{b}(1,2)}, \tag{2.1}$$

where indexes (1) and (2) refer to the first and second media, respectively. The field with the superscript "q" is a "forced" one, i.e., the charge field in the corresponding homogeneous infinite medium, and the field with the superscript "b" is a "free" one, i.e., the field arising due to the presence of the interface between two media (we use here the terminology introduced by V. L. Ginzburg [4]).

Note that the forced field was analyzed by us in [18]. Here we will focus only on obtaining the expression for it in the form that is convenient for solving the boundary problem under consideration.

Let us use the vector ($\vec{A}$) and scalar ($\Phi$) potentials connecting with the field components by the formulas $\vec{E} = -\frac{1}{c}\frac{\partial \vec{A}}{\partial t} - \nabla\Phi$, $\vec{B} = \mathrm{rot}\vec{A}$ (Gaussian system of units is used). We apply the Lorentz gauge, then the time Fourier transforms of the potentials obey the Helmholtz equation:

$$\left(\Delta + k_{1,2}^2\right) \begin{Bmatrix} \vec{A}_\omega^{(1,2)} \\ \Phi_\omega^{(1,2)} \end{Bmatrix} = -4\pi \begin{Bmatrix} \mu_{1,2} \vec{j}_\omega / c \\ \rho_{\Sigma\omega}/\varepsilon_{1,2} \end{Bmatrix}. \tag{2.2}$$

The forced field potentials $\vec{A}_\omega^{q(1,2)}$, $\Phi_\omega^{q(1,2)}$ also obey equations (2.2), and the free field potentials $\vec{A}_\omega^{b(1,2)}$, $\Phi_\omega^{b(1,2)}$ obey the same equations with zero right side.

We will solve equations (2.2) by the Fourier method. First of all, it is necessary to find the four-dimensional Fourier transforms of the charge and current densities. For the charge and current densities of the bunch itself, we have

$$\begin{Bmatrix} \rho_{\omega,\vec{k}} \\ j_{\omega,\vec{k}} \end{Bmatrix} = \frac{1}{(2\pi)^4} \begin{Bmatrix} 1 \\ v \end{Bmatrix} \iiint d^3R \cdot \delta(x,y) \delta(z-vt) e^{-i\vec{k}\vec{R}} \int q(t) e^{i\omega t} dt = \frac{1}{(2\pi)^3} \begin{Bmatrix} 1 \\ v \end{Bmatrix} q_\Omega, \tag{2.3}$$

where

$$q_\Omega = \frac{1}{2\pi} \int_{-\infty}^{\infty} q(t) e^{i\Omega t} dt, \qquad \Omega = \omega - vk_z. \tag{2.4}$$

Calculating the Fourier transform of the charge density of the "trace", we obtain

$$\rho_{1\,\omega,\vec{k}} = -\frac{\Omega}{(2\pi)^3 (\omega + i0)} q_\Omega, \tag{2.5}$$

The validity of this expression can be checked by calculating the inverse Fourier integral, which leads to (1.1). Summing up (2.3) and (2.5), for the total source we obtain

$$\rho_{\Sigma\omega,\vec{k}} = \frac{vk_z q_\Omega}{8\pi^3(\omega+i0)}, \quad \vec{j}_{\omega,\vec{k}} = \frac{vq_\Omega \vec{e}_z}{8\pi^3}. \tag{2.6}$$

In the case of bunch with the exponentially decreasing charge (1.2), for the Fourier transform of the bunch charge we have

$$q_\Omega = q_0\delta(\Omega) + \frac{q_0}{2\pi}\left[\frac{1}{i(\Omega+i0)} + \frac{\tau}{1-i\Omega\tau}\right]e^{i\Omega t_0}. \tag{2.7}$$

Note that the term «$+i0$» in the denominators (2.5) and (2.7) provides the required bypass of the pole.

Writing $\vec{A}_\omega^{q(1,2)}$, $\Phi_\omega^{q(1,2)}$ as the inverse Fourier integrals over the wave vector components, substituting them into equations (2.2) and equating the integrands, we obtain the fourfold Fourier images of the potentials. After that, we write the inverse Fourier transform over the wave vector components and pass to the cylindrical coordinate system both in the physical space ($r,\varphi,z$) and in the space of the wave vectors ($k_r, \varphi_k, k_z$). For the time Fourier transforms of the potentials, we obtain

$$\begin{Bmatrix} \vec{A}_\omega^{q(1,2)} \\ \Phi_\omega^{q(1,2)} \end{Bmatrix} = \frac{1}{2\pi^2}\int_{-\infty}^{\infty} dk_z \int_0^{\infty} dk_r \int_0^{2\pi} d\varphi_k \begin{Bmatrix} \mu_{1,2}\vec{\beta} \\ \dfrac{c\beta k_z}{(\omega+i0)\varepsilon_{1,2}} \end{Bmatrix} \frac{k_r q_\Omega \exp(ik_z z + ik_r r\cos(\varphi_k-\varphi))}{k_r^2+k_z^2-k_{1,2}^2}. \tag{2.8}$$

Since the integral over $\varphi_k$ reduces to the tabular one [21], Eq. (2.8) can be written as

$$\begin{Bmatrix} \vec{A}_\omega^{q(1,2)} \\ \Phi_\omega^{q(1,2)} \end{Bmatrix} = \frac{1}{\pi}\int_{-\infty}^{\infty} dk_z \int_0^{\infty} dk_r \begin{Bmatrix} \mu_{1,2}\vec{\beta} \\ \dfrac{c\beta k_z}{(\omega+i0)\varepsilon_{1,2}} \end{Bmatrix} \frac{k_r q_\Omega J_0(k_r r)\exp(ik_z z)}{k_r^2+k_z^2-k_{1,2}^2}, \tag{2.9}$$

where $J_0(k_r r)$ is Bessel function.

In contrast to [18], here it is convenient to represent the result as an integral over the tangential component $k_r$, therefore we take the integrals over $k_z$. Let us calculate the integral for the potential $\vec{A}$. Substituting (2.7) into (2.9) and calculating the integral with the delta function, we obtain

$$\vec{A}_\omega^{q(1,2)} = \frac{q_0\mu_{1,2}\vec{\beta}}{\pi}\int_0^\infty k_r J_0(k_r r)\left\{\frac{e^{i\omega z/v}}{v\left(k_r^2-s_{1,2}^2\right)} + \frac{e^{i\omega t_0}}{2\pi}\int_{-\infty}^{+\infty}\frac{e^{ik_z(z-vt_0)}}{k_z^2-\kappa_{1,2}^2}\left[\frac{1}{i(\Omega+i0)}+\frac{\tau}{1-i\Omega\tau}\right]dk_z\right\}dk_r, \tag{2.10}$$

where $s_{1,2} = \sqrt{\omega^2 v^{-2}\left(n_{1,2}^2\beta^2-1\right)}$, $\kappa_{1,2} = \sqrt{k_{1,2}^2-k_r^2}$. We set that these roots are defined by the rules $\operatorname{Im}s_{1,2} > 0$, $\operatorname{Im}\kappa_{1,2} > 0$. For the real parts, this leads to the rule $\operatorname{sgn}\operatorname{Re}s_{1,2} = \operatorname{sgn}\operatorname{Re}\kappa_{1,2} = \operatorname{sgn}\omega$.

The contour of integration over $k_z$ in (2.10) can be supplemented with the closed infinite semicircle located in the upper half-plane ($\operatorname{Im}k_z > 0$) at $z - vt_0 > 0$ and in the lower half-plane ($\operatorname{Im}k_z < 0$) at $z - vt_0 < 0$. After that, the integral over $k_z$ is found by calculating the residues at the poles $k_z = \kappa_{1,2}$, $\Omega = -i0$, $i\Omega\tau = 1$, if $z - vt_0 > 0$, and at the pole $k_z = -\kappa_{1,2}$, if $z - vt_0 < 0$. As a result, we obtain

$$\vec{A}_{\omega}^{q(1,2)}=\frac{q_0\mu_{1,2}\vec{\beta}}{\pi}\int_0^{\infty}k_rJ_0\left(k_rr\right)\left\{\frac{U_{1,2}}{2\omega\kappa_{1,2}}e^{i\omega t_0}e^{i\kappa_{1,2}|z-vt_0|}+\frac{e^{i\omega z/v}}{v}\left[\frac{\Theta\left(vt_0-z\right)}{k_r^2-s_{1,2}^2}+V_{1,2}\Theta\left(z-vt_0\right)\right]\right\}dk_r. \quad (2.11)$$

where

$$U_{1,2}=\frac{\omega}{\left[\omega-v\kappa_{1,2}\cdot\operatorname{sgn}\left(z-vt_0\right)\right]\left[1-i\tau\left(\omega-v\kappa_{1,2}\cdot\operatorname{sgn}\left(z-vt_0\right)\right)\right]},$$
$$V_{1,2}=\frac{\exp\left(-(z-vt_0)/(v\tau)\right)}{k_r^2-s_{1,2}^2-v^{-2}\tau^{-2}+2iv^{-2}\omega\tau^{-1}}. \quad (2.12)$$

Note that it is advisable to turn the integral along the semi axis into an integral along the entire real axis using the formula [22]

$$J_0(x)=\frac{H_0^{(1)}(x)-H_0^{(1)}(e^{i\pi}x)}{2}, \quad (2.13)$$

where $H_0^{(1)}(\xi)$ is Hankel function. Similarly, one can obtain the corresponding expression for $\Phi_{\omega}^{q(1,2)}$.

As a result, using the formulas $\vec{B}_{\omega}=\operatorname{rot}\vec{A}_{\omega}$, $\vec{E}_{\omega}=\frac{i\omega}{c}\vec{A}_{\omega}-\nabla\Phi_{\omega}$, we obtain the following expressions for the nonzero components of the forced electromagnetic field in each of the two media:

$$E_{r\omega}^{q(1,2)}=\frac{q_0\omega}{2\pi c^2\varepsilon_{1,2}}\int_{e^{i\pi}\infty}^{\infty}k_r^2H_1^{(1)}\left(k_rr\right)\left\{\operatorname{sgn}\left(z-vt_0\right)\frac{\beta U_{1,2}}{2k_0^3}e^{i\omega t_0}e^{i\kappa_{1,2}|z-vt_0|}+\right.$$
$$\left.+\frac{e^{i\omega z/v}}{k_0\beta}\left[\frac{\Theta\left(vt_0-z\right)}{k_r^2-s_{1,2}^2}+\left(1+\frac{i}{\omega\tau}\right)V_{1,2}\Theta\left(z-vt_0\right)\right]\right\}dk_r, \quad (2.14a)$$

$$E_{z\omega}^{q(1,2)}=\frac{iq_0\omega}{2\pi c^2}\int_{e^{i\pi}\infty}^{\infty}k_rH_0^{(1)}\left(k_rr\right)\left\{\frac{\beta k_r^2U_{1,2}}{2k_0^3\varepsilon_{1,2}\kappa_{1,2}}e^{i\omega t_0}e^{i\kappa_{1,2}|z-vt_0|}+\right.$$
$$\left.+e^{i\omega z/v}\left[\frac{n_{1,2}^2\beta^2-1}{\varepsilon_{1,2}\beta^2}\frac{\Theta\left(vt_0-z\right)}{k_r^2-s_{1,2}^2}+\left(\mu_{1,2}-\frac{\left(1+i\left(\omega\tau\right)^{-1}\right)^2}{\varepsilon_{1,2}\beta^2}\right)V_{1,2}\Theta\left(z-vt_0\right)\right]\right\}dk_r, \quad (2.14b)$$

$$H_{\varphi\omega}^{q(1,2)}=\frac{q_0\omega\beta}{2\pi c^2}\int_{e^{i\pi}\infty}^{\infty}k_r^2H_1^{(1)}\left(k_rr\right)\left\{\frac{U_{1,2}}{2k_0^2\kappa_{1,2}}e^{i\omega t_0}e^{i\kappa_{1,2}|z-vt_0|}+\right.$$
$$\left.+\frac{e^{i\omega z/v}}{k_0\beta}\left[\frac{\Theta\left(vt_0-z\right)}{k_r^2-s_{1,2}^2}+V_{1,2}\Theta\left(z-vt_0\right)\right]\right\}dk_r. \quad (2.14c)$$

## 3. Free field

The Fourier transform of free field component $E_{z\omega,\vec{k}_r}^{b(1,2)}\equiv E_{z\omega,k_x,k_y}^{b(1,2)}$ satisfies the equation

$$\frac{\partial^2}{\partial z^2}E_{z\omega,\vec{k}_r}^{b(1,2)}+\left(k_{1,2}^2-k_r^2\right)E_{z\omega,\vec{k}_r}^{b(1,2)}=0, \quad (3.1)$$

where $k_r^2=k_x^2+k_y^2$. Its solutions are the functions

$$E^{b(1,2)}_{z\omega,\vec{k}_r} = \frac{q_0\omega}{c^2} C_{\mp} e^{\mp i\kappa_{1,2} z}, \tag{3.2}$$

where the dimensional factor $q_0\omega/c^2$ is introduced for convenience. The signs "-" or "+" in (3.2) correspond to the waves propagating either in the negative (in medium 1) or in the positive (in medium 2) direction of $z$-axis. The Fourier transforms of other field components are easy to find using Maxwell equations.

Next, we write the corresponding inverse Fourier integrals of the form

$$F_\omega = \int_{-\infty}^{\infty}\int_{-\infty}^{\infty} F_{\omega,\vec{k}_r} e^{i\vec{k}_r\vec{r}} dk_x dk_y = \int_0^{\infty} dk_r \cdot k_r \int_0^{2\pi} d\varphi_k F_{\omega,\vec{k}_r} \exp\left(ik_r r\cos(\varphi_k - \varphi)\right). \tag{3.3}$$

After calculating the table integrals over $\varphi_k$ [21], the following representations for the nonzero field components are obtained:

$$\begin{Bmatrix} E^{b(1,2)}_{r\omega} \\ E^{b(1,2)}_{z\omega} \\ H^{b(1,2)}_{\varphi\omega} \end{Bmatrix} = 2\pi\frac{q_0\omega}{c^2}\int_0^{\infty} \begin{Bmatrix} \pm i\kappa_{1,2} J_1(k_r r) \\ k_r J_0(k_r r) \\ -ik_0\varepsilon_{1,2} J_1(k_r r) \end{Bmatrix} C_{\mp} e^{\mp i\kappa_{1,2} z} dk_r \tag{3.4}$$

Using (2.13) and similar relation for $J_1(x)$ [21], we write the field components as integrals over the entire real axis:

$$\begin{Bmatrix} E^{b(1,2)}_{r\omega} \\ E^{b(1,2)}_{z\omega} \\ H^{b(1,2)}_{\varphi\omega} \end{Bmatrix} = \pi\frac{q_0\omega}{c^2}\int_{e^{i\pi}\infty}^{\infty} \begin{Bmatrix} \pm i\kappa_{1,2} H^{(1)}_1(k_r r) \\ k_r H^{(1)}_0(k_r r) \\ -ik_0\varepsilon_{1,2} H^{(1)}_1(k_r r) \end{Bmatrix} C_{\mp} e^{\mp i\kappa_{1,2} z} dk_r. \tag{3.5}$$

To find the coefficients $C_{\mp}$, we use the usual boundary conditions

$$\begin{aligned} \varepsilon_1 E^{(1)}_{z\omega}\Big|_{z=0} &= \varepsilon_2 E^{(2)}_{z\omega}, \\ E^{(1)}_{r\omega}\Big|_{z=0} &= E^{(2)}_{r\omega}. \end{aligned} \tag{3.6}$$

They result in the system of algebraic equations which has the following solution:

$$C_+ = C_+^0 - \frac{i\beta k_r^2(\varepsilon_2\kappa_1 - \varepsilon_1\kappa_2)\tilde{U}_2}{4\pi^2 k_0^3 \varepsilon_2\kappa_2(\varepsilon_2\kappa_1 + \varepsilon_1\kappa_2)} e^{i(\omega+v\kappa_2)t_0}, \tag{3.7a}$$

$$C_- = C_-^0 - \frac{i\beta k_r^2 \tilde{U}_1}{4\pi^2 k_0^3 \varepsilon_1\kappa_1} e^{i(\omega+v\kappa_1)t_0} + \frac{i\beta k_r^2\tilde{U}_2}{2\pi^2 k_0^3(\varepsilon_1\kappa_2 + \varepsilon_2\kappa_1)} e^{i(\omega+v\kappa_2)t_0}, \tag{3.7b}$$

where

$$C_+^0 = \frac{i}{2\pi^2(\varepsilon_2\kappa_1 + \varepsilon_1\kappa_2)}\left[\left(\frac{n_1^2\beta^2 - 1}{\beta^2}\kappa_1 + \frac{k_r^2}{k_0\beta}\right)\frac{1}{k_r^2 - s_1^2} - \left(\frac{n_2^2\beta^2 - 1}{\beta^2}\kappa_1 + \frac{\varepsilon_1 k_r^2}{k_0\beta\varepsilon_2}\right)\frac{1}{k_r^2 - s_2^2}\right], \tag{3.8a}$$

$$C_{-}^{0}=-\frac{i}{2\pi^{2}\left(\varepsilon_{1}\kappa_{2}+\varepsilon_{2}\kappa_{1}\right)}\left[\left(\frac{n_{1}^{2}\beta^{2}-1}{\beta^{2}}\kappa_{2}-\frac{\varepsilon_{2}k_{r}^{2}}{k_{0}\beta\varepsilon_{1}}\right)\frac{1}{k_{r}^{2}-s_{1}^{2}}-\left(\frac{n_{2}^{2}\beta^{2}-1}{\beta^{2}}\kappa_{2}-\frac{k_{r}^{2}}{k_{0}\beta}\right)\frac{1}{k_{r}^{2}-s_{2}^{2}}\right], \quad (3.8b)$$

$$\tilde{U}_{1,2}=\frac{\omega}{\left(\omega+v\kappa_{1,2}\right)\left[1-i\tau\left(\omega+v\kappa_{1,2}\right)\right]}. \quad (3.9)$$

## 4. Radiation in medium 1

Further we investigate the field in the wave (far-field) area $|k_1|R\gg 1$. In this area, the main part of the electromagnetic field is the radiation field which has the greatest interest to us.

We assume that, in medium 1, the charge velocity does not exceed the phase velocity of the waves: $v<c/(\mathrm{Re}\,n_1)$. For the practically important case, when the medium 1 is close to vacuum, this condition is satisfied automatically.

Let us introduce the spherical coordinates $R$, $\theta$ ($r=R\sin\theta$, $z=R\cos\theta$), as well as the new integration variable $\chi$ such that $k_r=k_1\sin\chi$. The complete field, consisting of the forced and free fields, can be written as

$$\left\{\begin{matrix}E_{r\omega}^{(1)}\\ E_{z\omega}^{(1)}\\ H_{\varphi\omega}^{(1)}\end{matrix}\right\}=\pi\frac{q_0k_0^3n_1^2}{c}\int_{\Gamma}\left\{\begin{matrix}i\cos\chi H_1^{(1)}\left(k_1R\sin\theta\sin\chi\right)\\ \sin\chi H_0^{(1)}\left(k_1R\sin\theta\sin\chi\right)\\ -i\sqrt{\varepsilon_1/\mu_1}H_1^{(1)}\left(k_1R\sin\theta\sin\chi\right)\end{matrix}\right\}\tilde{C}_{-}\cos\chi\, e^{-ik_1R\cos\theta\cos\chi}d\chi, \quad (4.1)$$

where

$$\tilde{C}_{-}=C_{-}^{0}+\frac{i\beta k_r^2\tilde{U}_2}{2\pi^2k_0^3\left(\varepsilon_1\kappa_2+\varepsilon_2\kappa_1\right)}e^{i(\omega+v\kappa_2)t_0}. \quad (4.2)$$

The integration contour $\Gamma$ is shown in fig. 2.

For the approximate calculation of these integrals at $|k_1|R\gg 1$, one can use the steepest descent method. First, we need to transform the original integration contour $\Gamma$ to the fastest descent contour $\Gamma_*$ (Fig. 2). It can be shown that the poles of the integrand cannot be intersected during the contour transformation. However, the branch point $\chi=\chi_b$ determined by the formula $k_2=k_1\sin\chi_b$ can be intersected. In this case, the integral along the contour covering the cut appears. As it is known [20], the contribution of such a contour is a so-called “lateral” wave, which decreases with distance proportionally to $R^{-3/2}$. Further we will be interested only in the main term of the asymptotic which is proportional to $R^{-1}$. Therefore the contribution of the lateral wave can be neglected.

Using the asymptotic of the Hankel functions [22], we obtain from (4.1) the following approximate integrals over the fastest descent contour:

$$\left\{\begin{matrix}E_{r\omega}^{(1)}\\ E_{z\omega}^{(1)}\\ H_{\varphi\omega}^{(1)}\end{matrix}\right\}\approx\sqrt{2\pi}e^{-i\pi/4}\frac{q_0k_0^3n_1^2}{c}\int_{\Gamma_*}\frac{\tilde{C}_{-}\cos\chi}{\sqrt{k_1R\sin\chi\sin\theta}}\left\{\begin{matrix}\cos\chi\\ \sin\chi\\ -\sqrt{\varepsilon_1/\mu_1}\end{matrix}\right\}\exp\left\{ik_1R\cos\left(\chi-\left(\pi-\theta\right)\right)\right\}d\chi. \quad (4.3)$$

The saddle point of the integrand is located at $\chi = \chi_s = \pi - \theta$. The integrals (4.3) are calaulated with help of well-known formula of the steepest descent method [21]. Using the spherical coordinates $R,\ \theta,\ \varphi$, one can shown that the radiation field has only following components $E_\theta^{(1)}$ and $H_\varphi^{(1)}$:

$$E_{\theta\omega}^{(1)} \approx \sqrt{\frac{\mu_1}{\varepsilon_1}} H_{\varphi\omega}^{(1)} \approx -2\pi i \frac{q_0 k_0^2 n_1}{c \tan\theta} \frac{e^{ik_1 R}}{R} \tilde{C}_-\Big|_{\chi=\chi_s}. \tag{4.4}$$

Radiation emanating from a limited region of space is usually characterized by the spectral-angular density of the radiated energy. The expression for it is easy to obtain, based on the formula for the energy flux density of the spherical wave:

$$\vec{S} = \frac{c}{4\pi} E_\theta^{(1)} H_\varphi^{(1)} \vec{e}_R. \tag{4.5}$$

The total energy passing through the unit area orthogonal to this vector for the entire time is equal to

$$\int_{-\infty}^{\infty} S_R dt = \int_0^{\infty} c \left| \sqrt{\frac{\varepsilon_1}{\mu_1}} E_{\theta\omega}^{(1)} \right|^2 d\omega. \tag{4.6}$$

The integrand in (4.6), multiplied by $R^2$, is the spectral-angular radiation energy density:

$$\frac{d^2 W}{d\Omega d\omega} = cR^2 \left| \sqrt{\frac{\varepsilon_1}{\mu_1}} E_{\theta\omega}^{(1)} \right|^2 = \frac{q_0^2}{4\pi^2 c} |F|^2, \tag{4.7}$$

where

$$F = \frac{2\left(\varepsilon_1^3 \mu_1\right)^{1/4}}{\left(\varepsilon_1 N_{21} - \varepsilon_2 n_1 \cos\theta\right)\tan\theta} \left[ -\frac{\left(n_1^2 \beta^2 - 1\right) N_{21} - \varepsilon_2 \mu_1 \beta \sin^2\theta}{1 - n_1^2 \beta^2 \cos^2\theta} + \right.$$
$$\left. + \frac{\left(n_2^2 \beta^2 - 1\right) N_{21} - n_1^2 \beta \sin^2\theta}{1 - \beta^2 N_{21}^2} + \frac{n_1^2 \beta \sin^2\theta}{\left(1 + \beta N_{21}\right)\left[1 - i\omega\tau\left(1 + \beta N_{21}\right)\right]} e^{i\omega t_0 \left(1 + \beta N_{21}\right)} \right]. \tag{4.8}$$

Here we have introduced the notation $N_{21} = \sqrt{n_2^2 - n_1^2 \sin^2\theta}$.

In the case of homogeneous medium ($\varepsilon_2 = \varepsilon_1$, $\mu_2 = \mu_1$) the formula (4.8) goes to the following expression:

$$F = F_h = -\frac{\left(\varepsilon_1 \mu_1^3\right)^{1/4} \beta \sin\theta}{\left(1 - n_1 \beta \cos\theta\right)\left[1 - i\omega\tau\left(1 - n_1 \beta \cos\theta\right)\right]} e^{i\omega t_0 \left(1 - n_1 \beta \cos\theta\right)}. \tag{4.9}$$

This result coincides with one obtained in [20] if we put $t_0 = 0$.

In the nonrelativistic case ($\beta \ll 1$), keeping only quantities of order of $\beta$, we obtain

$$F = \frac{\left(\varepsilon_1^7 \mu_1^5\right)^{1/4} \beta \sin(2\theta)}{\varepsilon_1 N_{21} - \varepsilon_2 n_1 \cos\theta} \left[ \frac{\varepsilon_2 - \varepsilon_1}{\varepsilon_1} + \frac{1}{1 - i\omega\tau} e^{i\omega t_0 \left(1 + \beta N_{21}\right)} \right]. \tag{4.10}$$

We also note the special case when the second medium has a very large real permittivity or it is a good conductor (anyway $|\varepsilon_2| \gg |\varepsilon_1|$). Then it can be shown that

$$F \approx -\frac{2\left(\varepsilon_1 \mu_1^3\right)^{1/4} \beta \sin\theta}{1 - n_1^2 \beta^2 \cos^2\theta}. \tag{4.11}$$

This result coincides with one for the case of the unchanging charge flying into an ideal conductor (which was to be expected, since processes inside an ideal conductor do not affect the electromagnetic field outside it).

## 5. Radiation in medium 2

The asymptotics of the field in the medium 2 is calculated in a similar way. In this case, it is useful to introduce the new integration variable $\chi$, such that $k_r = k_2 \sin\chi$. Without going into details of the calculation, we note only its most important points.

Unlike medium 1, medium 2 may has some significant refractive index. Therefore it is possible that $|n_2|\beta > 1$. In such a situation, the pole determined by the equation $s_2 = k_2 \sin\chi_p$ (where $\chi_p = \mathrm{acos}\left(\left(n_2\beta\right)^{-1}\right)$) can be intersected during the transformation of the initial integration contour to the fastest descent path, and it can make some significant contribution. Physically, this means the presence of Cherenkov radiation (CR), i.e. the cylindrical wave diverging from the charge trajectory. This pole contributes to the forced field under condition $0 < z < vt_0 + r\cot\chi_p$, and to the free field under condition $0 < z < r\cot\chi_p$. However, it turns out that these contributions mutually compensate each other, so, in the total field, the pole contribution is present only in the region $r\cot\chi_p < z < vt_0 + r\cot\chi_p$ (Fig. 5). This region is the area where CR exists.

However, it should be noted that, on the boundaries of this region ( $z = r\cot\chi_p$ and $z = vt_0 + r\cot\chi_p$ ), the saddle point coincides with the pole, as a result of which the asymptotics presented below are incorrect. More precise ("uniform") asymptotics for the case of an infinite homogeneous medium are given in our paper [18]. These asymptotics are valid in the vicinity of the boundary $z = vt_0 + r\cot\chi_p$. The behavior of the field in the vicinity of the boundary $z = r\cot\chi_p$ is similar.

Spherical waves emanating from the point where the charge enters the medium 2 and from the region where the bunch loses the charge are determined by the contribution of the saddle point $\chi = \chi_s = \theta$.

Omitting calculations, we present the final result for the field asymptotics in medium 2. It can be written as

$$\vec{E} = \vec{E}^{(I)} + \vec{E}^{(II)}, \qquad \vec{H} = \vec{H}^{(I)} + \vec{H}^{(II)}. \tag{5.1}$$

Here, the term with the index $(I)$ is the contribution of the saddle point. In the spherical coordinates, it has only components $E_\theta^{(I)}$ and $H_\varphi^{(I)}$:

$$E_{\theta\omega}^{(I)} \approx \sqrt{\frac{\mu_2}{\varepsilon_2}} H_{\varphi\omega}^{(I)} \approx \frac{q_0}{c}\left[\frac{2\pi i k_0^2 n_2}{\tan\theta} C_+\Big|_{\chi=\theta} - \frac{\beta\mu_2}{2\pi}\sin\theta\, U_2\Big|_{\chi=\theta} e^{i\omega(1-n_2\beta\cos\theta)t_0}\right]\frac{e^{ik_2R}}{R}, \tag{5.2}$$

The term with the index $(II)$ is the contribution of the pole. If $n_2\beta < 1$ then it is insignificant (exponentially decreases with increasing in $r$). If $n_2\beta > 1$ then it is a cylindrical wave of CR with components

$$\begin{Bmatrix} E_{r\omega}^{(II)} \\ E_{z\omega}^{(II)} \\ H_{\varphi\omega}^{(II)} \end{Bmatrix} = \frac{q_0}{\sqrt{2\pi c}} \begin{Bmatrix} s_2/(\beta\varepsilon_2) \\ -cs_2^2/(\varepsilon_2\omega) \\ s_2 \end{Bmatrix} \frac{\exp\left(i\omega z/v + is_2 r - i\pi/4\right)}{\sqrt{s_2 r}} \quad \text{for } r\cot\chi_p < z < vt_0 + r\cot\chi_p, \tag{5.3}$$

This radiation exists only in the band indicated in (5.3) (see Fig. 3).

The necessary condition for the applicability of the asymptotics (5.1) - (5.3) is the inequality $|k_2 R| >> 1$ that determines the "wave" zone. In the case $n_2\beta > 1$, the asymptotics is applicable only if the saddle point $\chi = \theta$ is sufficiently far from the pole $\chi = \chi_p$. This condition is reduced to the requirement that the observation point is not close to the boundaries of the region of existence of a cylindrical wave: $k_2 \left| z\tan\chi_p - r \right| >> 1$, $k_2 \left| (z - vt_0)\tan\chi_p - r \right| >> 1$. In the region $r\cot\chi_p < z < vt_0 + r\cot\chi_p$, the cylindrical wave of CR (5.3), which decreases proportionally to $1/\sqrt{r}$, dominates the spherical wave, which decreases proportionally to $1/R$.

The spectral-angular density of energy of the spherical wave (5.2) is determined by the expression

$$\frac{d^2W_2}{d\Omega d\omega} = cR^2 \left| \sqrt{\frac{\varepsilon_2}{\mu_2}} E_{\theta\omega}^{(2)} \right|^2 = \frac{q_0^2}{4\pi^2 c} |F|^2, \tag{5.4}$$

where

$$F = \frac{2\left(\varepsilon_2^3\mu_2\right)^{1/4}}{\left(\varepsilon_2 N_{12} + \varepsilon_1 n_2\cos\theta\right)\tan\theta} \left[ -\frac{\left(n_1^2\beta^2 - 1\right)N_{12} + n_2^2\beta\sin^2\theta}{1 - \beta^2 N_{12}^2} + \frac{\left(n_2^2\beta^2 - 1\right)N_{12} + \varepsilon_1\mu_2\beta\sin^2\theta}{1 - \beta^2 n_2^2\cos^2\theta} \right] - \\ -\beta\left(\varepsilon_2\mu_2^3\right)^{1/4}\sin\theta \left[ W_- e^{-i\omega t_0 n_2\beta\cos\theta} - W_+ \frac{\varepsilon_2 N_{12} - \varepsilon_1 n_2\cos\theta}{\varepsilon_2 N_{12} + \varepsilon_1 n_2\cos\theta} e^{i\omega t_0 n_2\beta\cos\theta} \right] e^{i\omega t_0}. \tag{5.5}$$

Here $N_{12} = \sqrt{n_1^2 - n_2^2\sin^2\theta}$, $W_\pm = \dfrac{1}{\left(1 \pm n_2\beta\cos\theta\right)\left[1 - i\tau\omega\left(1 \pm n_2\beta\cos\theta\right)\right]}$. For nonrelativistic motion of charge, simplifying (5.5), one can obtain the following approximate expression:

$$F = \frac{\beta\left(\varepsilon_2^7\mu_2^5\right)^{1/4}\sin(2\theta)}{\varepsilon_2 N_{12} + \varepsilon_1 n_2\cos\theta}\frac{\varepsilon_1 - \varepsilon_2}{\varepsilon_2} - \\ -\frac{\beta\left(\varepsilon_2\mu_2^3\right)^{1/4}}{1 - i\omega\tau}\sin\theta\left( e^{-i\omega t_0 n_2\beta\cos\theta} - \frac{\varepsilon_2 N_{12} - \varepsilon_1 n_2\cos\theta}{\varepsilon_2 N_{12} + \varepsilon_1 n_2\cos\theta} e^{i\omega t_0 n_2\beta\cos\theta} \right) e^{i\omega t_0}. \tag{5.6}$$

## 6. Numerical results and discussion

The results of calculation of the spectral-angular density of the spherical wave energy depending on the angle $\tilde{\theta} = \pi - \theta$ (see Fig.1) for various values of $\beta$, $\omega t_0$ and $\omega\tau$ are shown in Figs. 4–6. The first medium (the region $0 < \tilde{\theta} < 90^0$) is assumed to be vacuum, and the second medium (the region $90^0 < \tilde{\theta} < 180^0$) characterized by the constants $\varepsilon_2 = 1.5$, $\mu_2 = 1$.

Recall that, in accordance with (4.11), in the unbounded homogeneous medium, radiation is emitted mainly "forward" [18]. The maximum of the angular density is achieved at some acute angle with respect to the motion direction. As the velocity $\beta$ increases, the maximum increases and tends to $\theta = 0$ if $\beta \to 1$.

In the presence of the interface, the angular dependence changes radically. First, we note that the radiation in the second medium is more intense than in the first one (in Figs. 4–6, the different scales on the vertical axis are used for regions 1 and 2). This difference increases with increasing the velocity. For ultra relativistic bunches, the difference between the main maximums in regions 1 and 2 can be several orders of magnitude (Fig. 6).

Note that the radiation does not propagate along $z$ -axis and along the interface ($\theta = \pi / 2$). The last fact is explained by that the wave incidences on the boundary at the "sliding" angle; in this case, the reflection coefficient is equal to $-1$, and the total field is equal to zero.

Let us consider in detail Fig. 4, illustrating the case of the relatively low velocity ($\beta = 0.1$). If the process of the bunch decay starts directly from the interface ($t_0 = 0$), then both in the first (vacuum-like) region ($0 < \tilde{\theta} < 90^0$ ) and in the second region ($90^0 < \tilde{\theta} < 180^0$) there is only one maximum (Fig. 4, top). Both maximums have a relatively large width and decrease with increasing the bunch decay time $\tau$ . The latter fact is explained by that the number of particles stopping per unit time decreases, which leads to the decrease of the corresponding part of the spherical wave.

Note that, in the region $90^0 < \tilde{\theta} < 180^0$ , the maximum has a character of a "fracture". It is observed when the angle is equal to the limiting angle of the total internal reflection: $\theta = \chi_b = \arcsin\left(n_1/n_2\right) \approx 55^0$ . Mathematically, this fracture is related to the fact that at such an angle the saddle point of the integrand coincides with its branch point. Strictly speaking, the obtained asymptotics is not valid at $\left|\theta - \chi_b\right| \le 1/\left(k_2 R\right)$, but, due to the condition $k_2 R >> 1$, this angular region is insignificant.

If the process of the bunch decay begins at some significant distance from the interface, then the complex dependency consisting from several maximums and minimums can take place in both regions (Fig. 4, bottom). This effect is explained by the fact that the waves emmited from the different zones are added: from the point of entry of the charge into medium 2 and from the zone of the bunch decay (in the second medium there is also the wave reflected from the boundary). This effect is observed if the waves amplitudes are comparable in magnitude. If the bunch decay is sufficiently slow, then the role of the corresponding spherical wave is insignificant, and the effect noted above disappears: the radiation is close to the ordinary transition radiation (the dotted curves).

In the case of $\beta = 0.5$ (Fig. 5), the regularities noted above do not qualitatively change. We note only that, with increase in the velocity, the general increase in the radiation energy occurs, and the number of extrema at $\omega t_0 >> 1$ increases also (cf. Fig. 5 and 4).

Figure 6 illustrates the case when the charge velocity is higher than the wave velocity in the second medium ($n_2 \beta > 1$), and, correspondingly, Cherenkov radiation is generated. It should be emphasized that CR itself is not shown in these graphs. Its field decreases proportionally to $1/\sqrt{r}$ , therefore it is much larger than the field of spherical waves. However, it exists only in the area $\left|\theta - \chi_p\right| \le v t_0 / R$, which is small if $R >> v t_0$ (note that $\chi_p \approx 34^0$ for Fig. 6 ). The graphs are not correct in this narrow angular region, since the saddle point coincides with the pole at $\theta = \chi_p$ (uniform asymptotics, which is correct at $\theta \approx \chi_p$, for the case of a homogeneous infinite medium, was obtained in [18]).

In the case $n_2\beta > 1$, the following features of the spherical wave can be noted. First, it is much more intense than in the case $n_2\beta < 1$, which can be seen from comparison of Fig. 6 with Figs. 3 and 4. In the vacuum region, the main maximum shifts towards the larger angles with increasing the velocity, and it is close to $180^0$ at $\beta \approx 1$ (but never reaches this value). In the region $0 < \theta < 90^0$, if the values $\omega t_0$ and/or $\omega\tau$ are sufficiently large, then there is the sharp maximum at the angle $\theta = \chi_p$. Thus, the transition from the "subluminal" ($n_2\beta < 1$) mode of the charge motion to the "superluminal" ($n_2\beta > 1$) mode leads not only to the appearance of Cherenkov radiation, but also to radical changes of the spherical wave properties.

## Conclusion

We studied the electromagnetic radiation of the charged bunch of small size crossing the flat interface between two media. It was assumed that the charge magnitude decreases exponentially starting from some time moment after the charge enters the second medium. It was taken into account that the filamentous "trace" consisting of immobile charges is formed in the second medium. It was assumed that Cherenkov radiation could be generated in the second medium only.

It was obtained the general problem solution which is the sum of the forced field, i.e. the charge field in the unbounded medium, and the free field due to the influence of the interface. The asymptotic study for the wave zone was carried out. We have found the expressions for the spherical wave as well for the cylindrical wave generated in the second medium at the sufficiently high charge velocity. The spherical wave is radically different from the usual transition radiation, since it consists of two parts: the transition radiation wave and the wave generated due to the bunch decay and the formation of its "trace".

The series of computations was performed for the case when the charge flies from vacuum into the optically denser medium. If the process of the bunch decay starts from the interface, then the angular distribution of the radiation energy has one maximum in each from two regions. In the second medium, the radiation, as a rule, is greater than in the vacuum area, and this difference increases with increasing the charge velocity. If the charge velocity is higher than the speed of light in the second medium, then, along with the appearance of Cherenkov radiation, the properties of the spherical wave change radically. In particular, the main maximum of the angular distribution of the radiation energy increases sharply. If the distance from the charge entry point to the region of the bunch decay is sufficiently large, then the complex interference pattern with many extrema arises, due to the summation of different spherical waves.

## Acknowledgments

The study was supported by the Russian Science Foundation (project No. 18-72-10137) and the China Scholarship Programs.

## References

1. I.M. Frank. Vavilov-Cherenkov radiation: theoretical aspects (Nauka, Moscow, 1988; in Russian).
2. J.V. Jelley. Cerenkov radiation and its applications (Pergamon Press, 1958).
3. V.P. Zrelov. Vavilov-Cherenkov radiation in high-energy physics (Israel Program for Scientific Translations, Jerusalem, 1970).
4. V.L. Ginzburg, V.N. Tsytovich. Transition radiation and transition scattering (Hilger, London, 1990).
5. G.N. Afanasiev. Vavilov-Cherenkov and synchrotron radiation: foundations and applications. (Dordrecht: Kluwer Academic Publishers, 2004).
6. B.M. Bolotovskii. Usp. Fiz. Nauk, Theory of Vavilov-Cherenkov effect, v. 62 (3), 201 (1957) (in Russian).
7. B.M. Bolotovskii, Theory of Cherenkov radiation (III), Physics–Uspekhi, v. 4, 781 (1962).
8. S.N. Galyamin,A.V. Tyukhtin, Electromagnetic field of a charge travelling into an anisotropic medium, Phys. Rev. E, v. 84, 056608 (2011).
9. S.N. Galyamin, D.Ya. Kapshtan, A.V. Tyukhtin, Electromagnetic field of a charge moving in a cold magnetized plasma, Phys. Rev. E, v. 87, 013109 (2013).
10. M. Marongiu, E. Chiadroni, M. Croia, M. Ferrario, L. Ficcadenti, S. Lupi1, V. Martinelli, A. Mostacci1, R. Pompili, S. Tofani, Electromagnetic and beam dynamics studies for high gradient accelerators at terahertz frequencies, J. of Physics: Conf. Ser., v. 1596, 012029 (2020).
11. S.S. Baturin, A. Zholents, Stability condition for the drive bunch in a collinear wakefield accelerator, Phys. Rev. Accel. Beams, v. 21, 031301 (2018).
12. V.I. Boiko, V.A. Skvortsov, V.E. Fortov, I.V. Shamanin. Interaction of pulsed beams of charged particles with matter (Moscow, Fizmatlit, 2003; in Russian).
13. A.P. Chernyaev, A.V. Belousov, E.N. Lykova. Interaction of ionizing radiation with matter (Moscow, MGU, 2019; in Russian).
14. M.A. Baturitsky. Interaction of ionizing radiation with matter (Minsk, MGEU, 2005; in Russian).
15. S.H. Park, J.O. Kang, Basics of particle therapy I: physics, Radiation Oncology Journal, v. 29, 135 (2011).
16. K.W. Jang, W.J. Yoo, S.H. Shin, D. Shin, B. Lee, Fiber-optic Cerenkov radiation sensor for proton therapy dosimetry, Optics Express, v. 20(13), 13907 (2012).
17. S.G. Stuchebrov, Yu.M. Cherepennikov, A.A. Krasnykh, I.A. Miloichikova, A.V. Vukolov, The method for the electron beam cross section measurement based on the detection of Cherenkov radiation in dielectric fiber, J. of Instrumentation, v. 13, C05020 (2018).
18. A.V. Tyukhtin, X. Fan, Radiation from a moving bunch of particles with a variable charge, Technical Physics, v. 67, 1222 (2022).
19. S.N. Galyamin, A.V. Tyukhtin, A. Kanareykin, P. Schoessow, Reversed Cherenkov-transition radiation by a charge crossing a left-handed medium boundary, Phys. Rev. Lett., v. 103, 194802 (2009).
20. S.N. Galyamin, A.V. Tyukhtin Electromagnetic field of a moving charge in the presence of a left handed medium, Phys. Rev. B, v. 81, 235134 (2010).
21. A.P. Prudnikov, Y.A. Brychkov, O. I. Marichev, Integrals and series, v. 1: Elementary functions (Gordon & Breach Science Publications, New York, 1986).
22. M. Abramowitz, I. A. Stegun, Handbook of mathematical functions: with formulas, graphs, and mathematical tables (Dover Publications, New York, 1972).

23. L.M. Brekhovskih, Waves in layered media (Academic Press, New York, 1980).
24. L. Felsen, N. Marcuvitz. Radiation and scattering of waves (Wiley Interscience, New York, 2003).

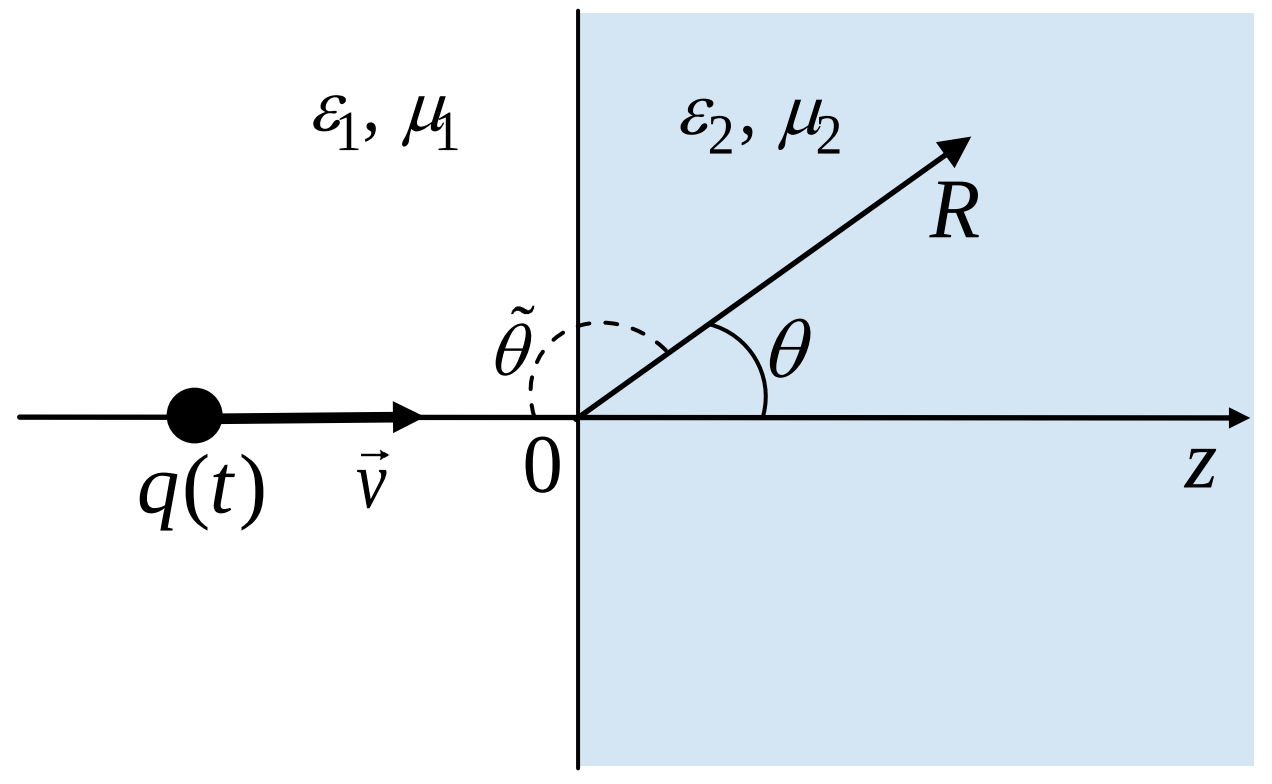

Fig.1. The problem geometry.

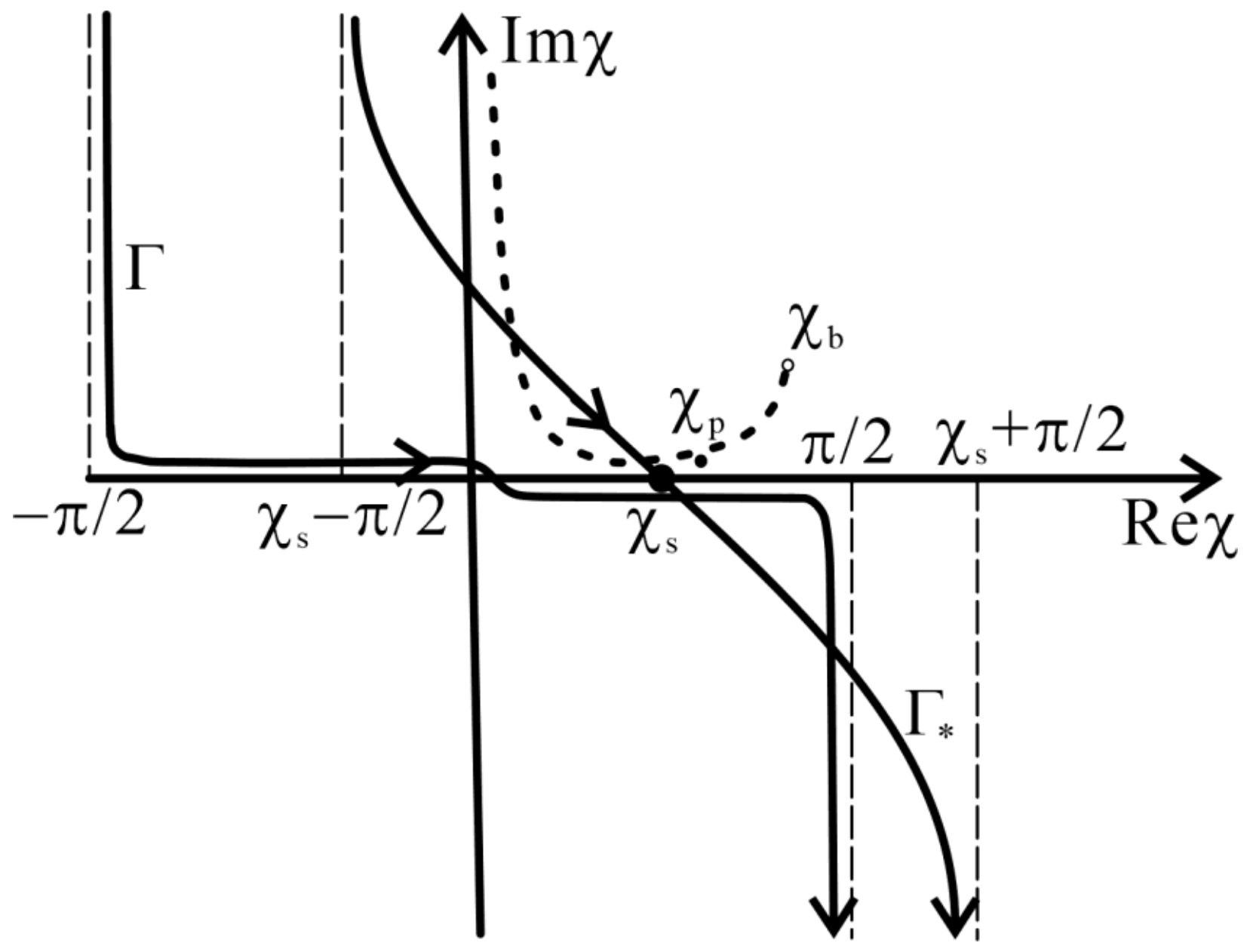

Fig. 2. The initial contour of integration $\Gamma$ and the contour of the fastest descent $\Gamma_*$ on the complex plane $\chi$; the dotted line shows the cut.

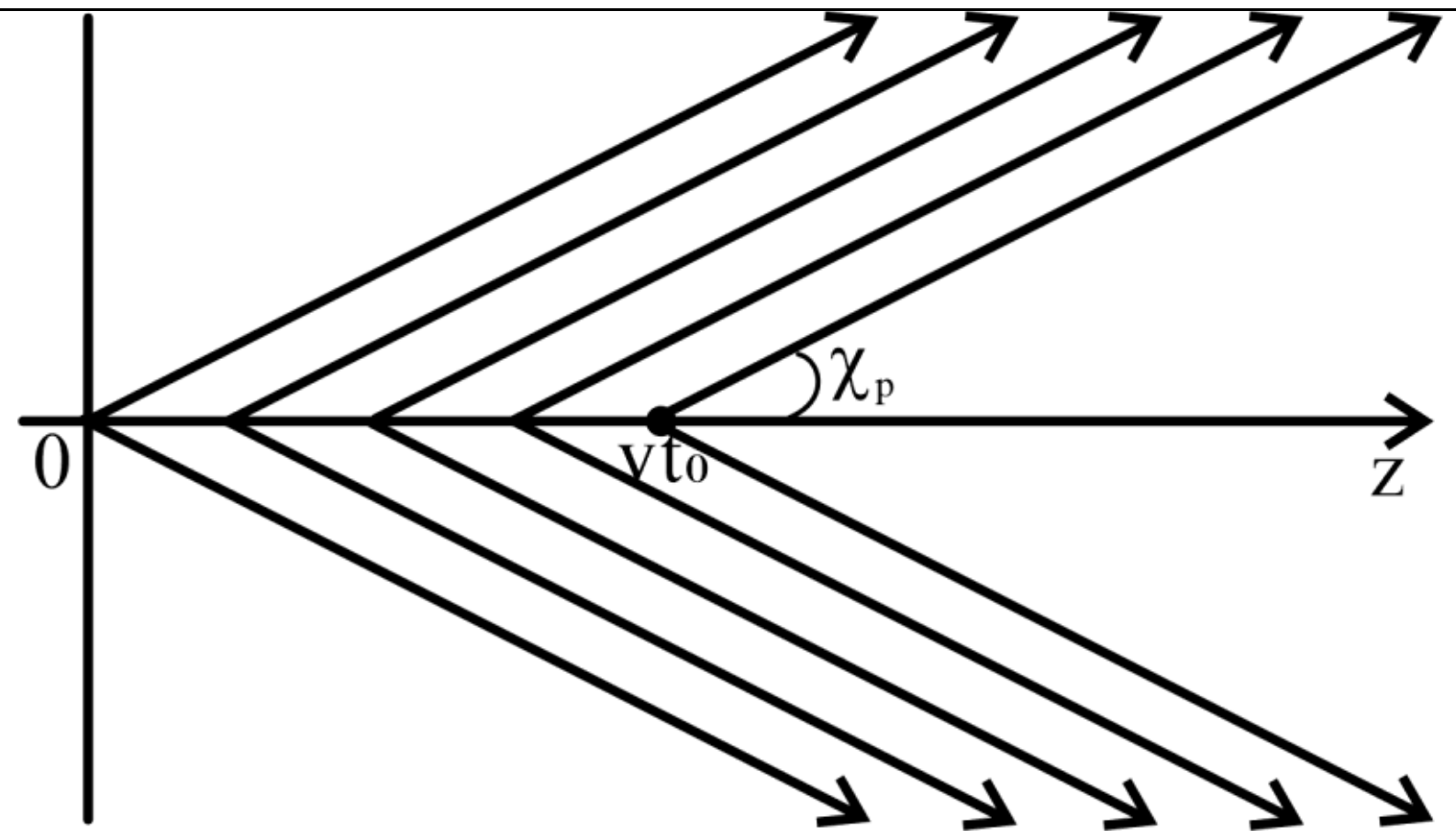

Fig. 3. The area of Cherenkov radiation.

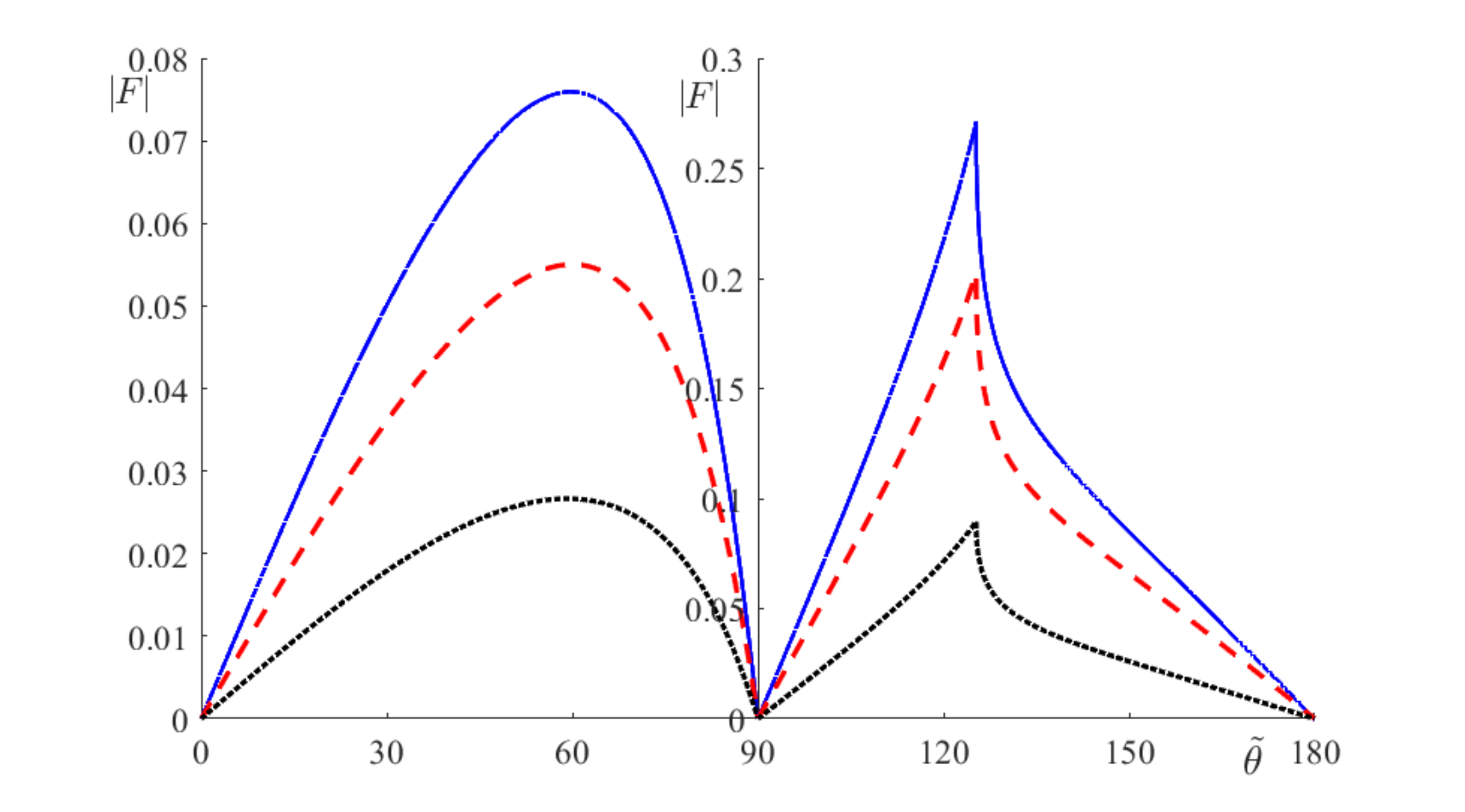

Fig. 4. Dependence of $|F|$ on the angle $\tilde{\theta} = \pi - \theta$ in the case $\beta = 0.1$. Other parameters: $\varepsilon_1 = \mu_1 = \mu_2 = 1$, $\varepsilon_2 = 1.5$; $\omega\tau = 0$ (blue solid curve), $\omega\tau = 1$ (red dashed curve), $\omega\tau = 50$ (black dotted curve); for the high plot $\omega t_0 = 0$, for the low plot $\omega t_0 = 100$.

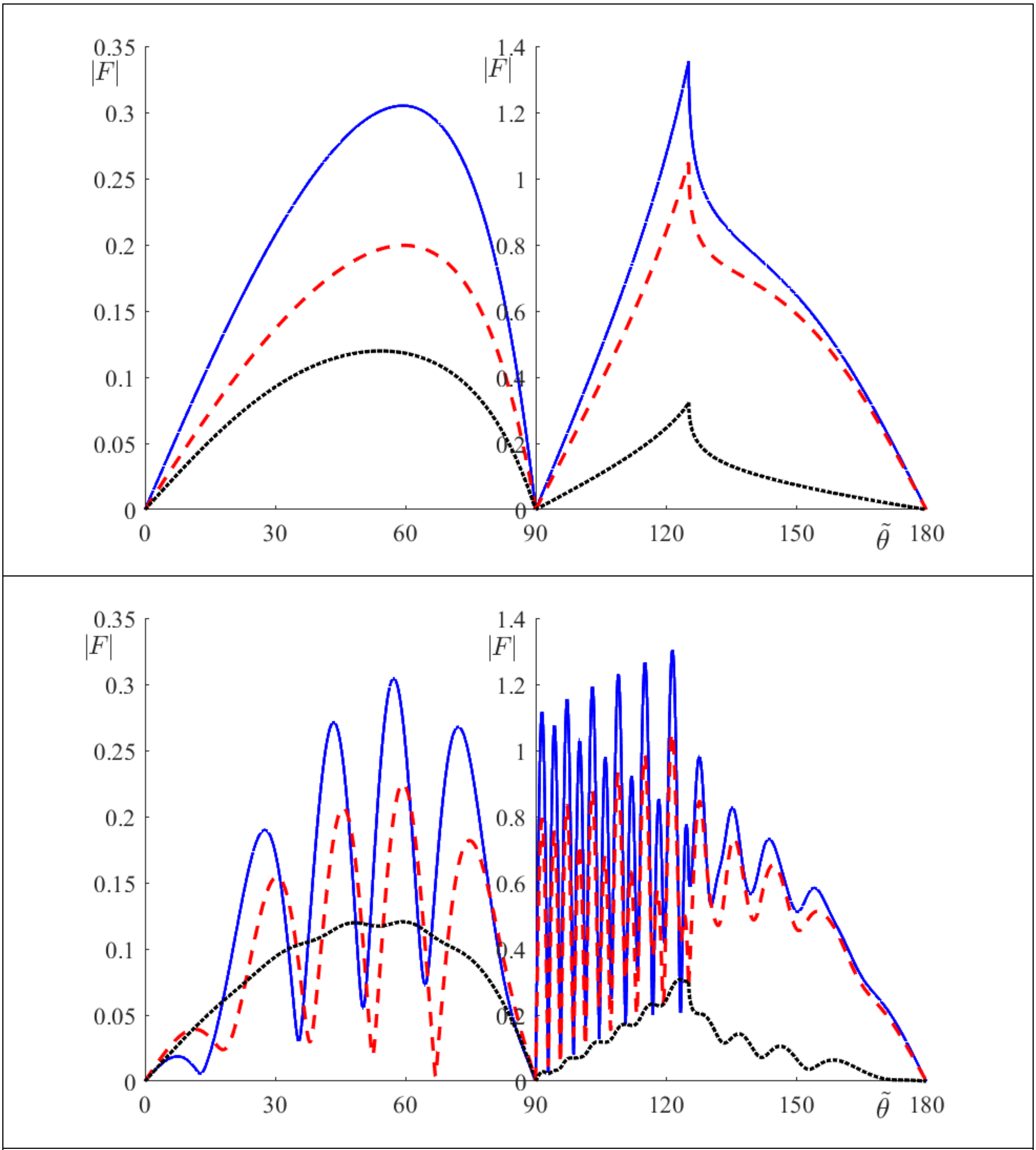

Fig. 5. The same as in Fig. 4 for $\beta = 0.5$.

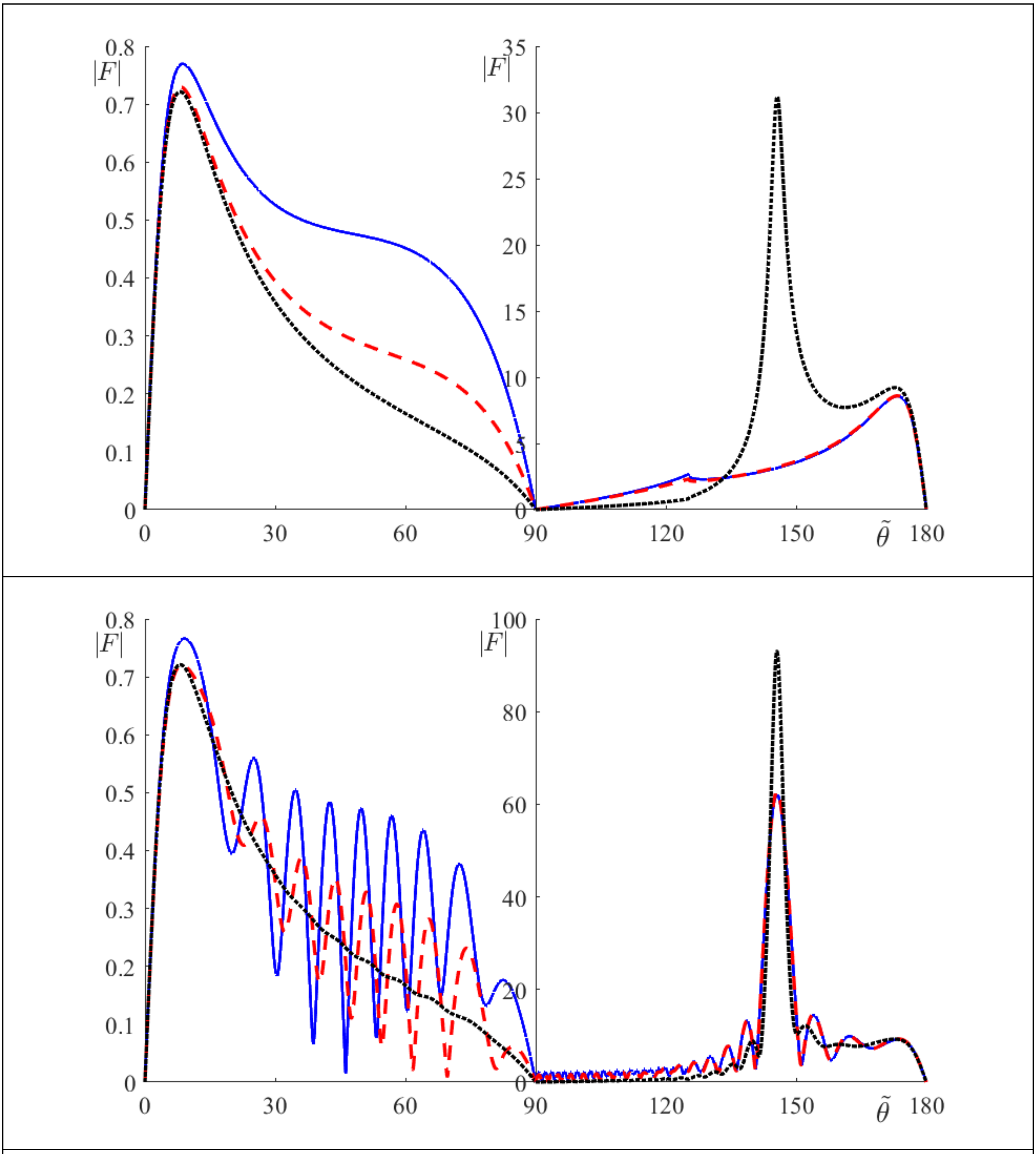

Fig. 6. The same as in Fig. 4 for $\beta = 0.99$.